**Processing Induced Distinct Charge Carrier Dynamics of Bulky Organic Halide Treated Perovskites**

Benjia Dak Dou [a, b], Dane W. deQuilettes [a, b], Madeleine Laitz [a, b], Roberto Brenes [a,b], Lili Wang [c], Ella L Wassweiler [a,b], Richard Swartwout [a,b], Jason J. Yoo [c, d], Melany Sponseller [a,b], Noor Titan Putri Hartono [e], Shijing Sun [e], Tonio Buonassisi [e], Moungi G Bawendi [c], Vladimir Bulović [a,b,*]

[a] Research Laboratory of Electronics, Massachusetts Institute of Technology, 77 Massachusetts Avenue, Cambridge, Massachusetts 02139, USA. bulovic@mit.edu
[b] Department of Electrical Engineering and Computer Science, Massachusetts Institute of Technology, 77 Massachusetts Avenue, Cambridge, Massachusetts 02139, USA.
[c] Department of Chemistry, Massachusetts Institute of Technology, 77 Massachusetts Avenue, Cambridge, Massachusetts 02139, USA.
[d] Division of Advanced Materials, Korea Research Institute of Chemical Technology, Daejeon, Republic of Korea
[e] Department of Mechanical Engineering, Massachusetts Institute of Technology, 77 Massachusetts Avenue, Cambridge, Massachusetts 02139, USA

TOC

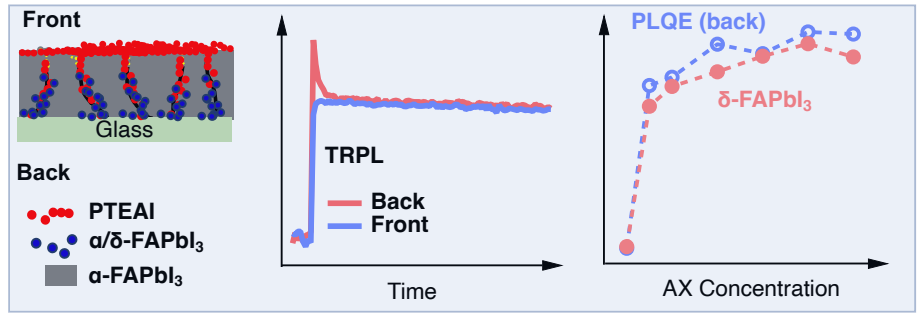

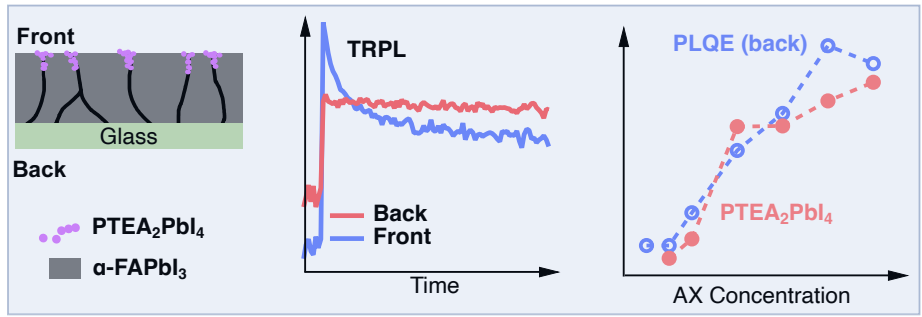

State-of-the-art metal halide perovskite-based photovoltaics often employ organic ammonium salts, AX, as a surface passivator, where A is a large organic cation and X is a halide. These surface treatments passivate the perovskite by forming layered perovskites (e.g., $A_2PbX_4$) or by AX itself serving as a surface passivation agent on the perovskite photoactive film. It remains unclear whether layered perovskites or AX is the ideal passivator due to an incomplete understanding of the interfacial impact and resulting photoexcited carrier dynamics of AX treatment. In the present study, we use time resolved photoluminescence (TRPL) measurements to selectively probe the different interfaces of glass/perovskite/AX to demonstrate the vastly distinct interfacial photoexcited state dynamics with the presence of $A_2PbX_4$ or AX. Coupling the TRPL results with X-ray diffraction measurements and nanoscale microscopy measurements, we find that the presence of AX not only passivates the traps at the surface and the grain boundaries, but also induces an α/δ-FAPbI$_3$ phase mixing that alters the carrier dynamics near the glass/perovskite interface and enhances the photoluminescence quantum yield. In contrast, the passivation with $A_2PbI_4$ is mostly localized to the surface and grain boundaries near the top surface where the availability of $PbI_2$ directly determines the formation of $A_2PbI_4$. Such distinct mechanisms significantly impact the corresponding solar cell performance, and we find AX passivation that has not been converted to a layered perovskite allows for a much larger processing window (e.g., larger allowed variance of AX concentration which is critical for improving the eventual manufacturing yield) and more reproducible condition to realize device performance improvements, while $A_2PbI_4$ as a passivator yields a much narrower processing window. We expect these results to not only enable rapid optimization of existing perovskite/AX interfaces, but also to lead to more advanced AX treatments with phase compositions predesigned for highly efficient optoelectronic devices.

Metal halide perovskites (MHP) are an exceptional family of semiconductors for optoelectronic devices, exhibiting high absorption coefficients across the visible spectrum, long carrier lifetimes, high defect tolerance and facile bandgap tunability.[1,2] Combining these properties with high throughput roll-to-roll solution printability[3–8] makes MHPs one of the most promising semiconductors for high-performance, low-cost optoelectronic devices. In particular, MHP-based photovoltaics have now reached 25.5% certified power conversion efficiency (PCE), and 29.5% PCE when coupled with silicon cells in a tandem format.[9] The rapid advances in PCE have generated great interest in MHPs as a next-generation photovoltaic (PV) active layer.[10] Key discoveries that enable such rapid progress include new device architectures[11–13], MHP compositional engineering[14–17], processing solvent engineering,[18,19] and, most recently, defect passivation strategies.[20–22] Defect passivation has been most successfully accomplished via the incorporation of bulky organic halides AX (A: R-Ammonium, X: halide) as the passivating agent, resulting in multiple record-setting solar cells.[9,21–24]

However, it is still hotly debated what AX processing conditions (e.g., concentration, annealing temperature, etc.) lead to optimized device performance as a detailed mechanistic understanding of how AX treatments function over a wide parameter space is still missing. For example, You *et al.*[23] report that AX treatments employing phenethylammonuim iodide should not be annealed post-treatment for the best device performance. They argue that AX passivates the surface defects better than its layered perovskite counterpart, which is formed upon annealing from the AX precursor. Consistent with this perspective, Liu *et al*[25] recently designed a thermally stable AX treatment (e.g., (phenylene)di(ethylammonium) iodide) to prevent formation of a layered perovskite, and report enhanced device performance. In contrast, other reports suggest that forming the layered perovskite or 2-dimensional (2D) perovskite to create a 2D/3D perovskite heterointerface is key to effectively passivating defects in perovskites.[24,26,27] Such a concept of forming 2D/3D interface is, as Reese *et al.* argues[28], universal across different polycrystalline thin-film semiconductors (e.g., CdTe, CIGS, perovskite). In addition to this lack of consensus in the field, attempting to decipher the optimal concentration and processing conditions of AX through empirical device optimization is difficult, and there are reports of optimal AX treatment concentrations spanning over an order of magnitude, ranging from 1 mM to 40 mM.[21,23,25–27,29–31] Overall, these unknowns make the optimization of performance-enhancing AX treatments time-consuming, nonsystematic, and, more importantly, unscalable. These inconsistencies also highlight the challenges of decoupling the impact of AX material selection (i.e., intrinsic properties of the materials) versus processing conditions when developing AX passivation strategies for MHP. To better understand these contradictory reports and utilize AX treatments to their full potential, a systematic understanding of the carrier dynamics of perovskite/AX interaction is critical. Particularly, while some report a fast carrier lifetime when the film is excited from the AX treated side,[27,30] discussions of the interfacial conditions that sustain such carrier dynamics is absent.

In the present study, using formamidinium lead iodide ($FAPbI_3$) and phenyltriethylammonium iodide (PTEAI) as a model AX material, we probe the bulk interface dynamics of $FAPbI_3$/PTEAI treated films through time-resolved photoluminescence (TRPL) lifetime analyses. $FAPbI_3$ was chosen as it possesses excellent thermal stability and the desired bandgap for single junction cells,[32,33] and PTEAI was chosen as it improves the thermal and humidity stability of perovskites compared to dozens of other AX perovskite capping layer candidates.[34] To better understand how changes in structure and composition impact carrier dynamics, we deploy PL spectroscopy/microscopy, X-ray diffraction (XRD), atomic force microscopy (AFM), and scanning electron microscopy (SEM) to provide a comprehensive picture of $FAPbI_3$/PTEAI treated structures. We demonstrate distinct photoexcited carrier dynamics at the interface of glass/FAPbI$_3$/PTEAI with respect to the formation of PTEAI or PTEA$_2$PbI$_4$. While both PTEAI and PTEA$_2$PbI$_4$ passivate traps, their impact on the interfaces of glass/FAPbI$_3$/PTEAI treated samples is quite different. The presence of PTEAI not only alters perovskite surfaces, but also induces an α/δ-FAPbI$_3$ phase mix near the glass/perovskite interface that significantly enhances the photoluminescence of the film. In comparison, PTEA$_2$PbI$_4$ mainly forms at the perovskite top surface and near-surface grain boundaries of the films and thus nearly exclusively impacts the perovskite top surface. We reveal the processing conditions and transition points for forming PTEAI and PTEA$_2$PbI$_4$ on FAPbI$_3$ and its implications for device performance improvements. While PTEAI offers a much larger processing window (e.g., 10 - 60 mM) and higher reproducibility for improving device performance, the processing window yielding enhanced device performance for PTEA$_2$PbI$_4$ is much smaller (up to 10 mM), and improvements are highly dependent on the position and availability of PbI$_2$. This detailed description of photoexcited carrier dynamics at the PTEAI/FAPbI$_3$ interface reveals important physical insights that allow for the effective utilization of AX passivation across different perovskite compositions and processing methods.

**Film orientation-dependent carrier dynamics in PTEAI treated FAPbI$_3$/glass**
To create a more holistic picture of how FAPbI$_3$/PTEAI treatments impact the charge carrier dynamics under various processing conditions, we performed TRPL measurements on the samples. We note that the TRPL signal of most thin films is a combination of both bulk and surface recombination effects and, therefore, can be difficult to deconvolve these competing processes. However, for polycrystalline perovskite thin films, surface recombination is often the dominant recombination,[35–37] and it is assumed that the fast decay component ($\tau_1$) in the TRPL trace is a reflection of surface/interface-related carrier dynamics; whereas the slow decay component ($\tau_2$) is a reflection of bulk carrier recombination.[38] Under this assumption, TRPL measurements were performed from both the back side (glass substrate) and front side (perovskite film side), as shown in Fig. 1a, with the aim of understanding the impact of PTEAI on the surface and interface of FAPbI$_3$/glass.

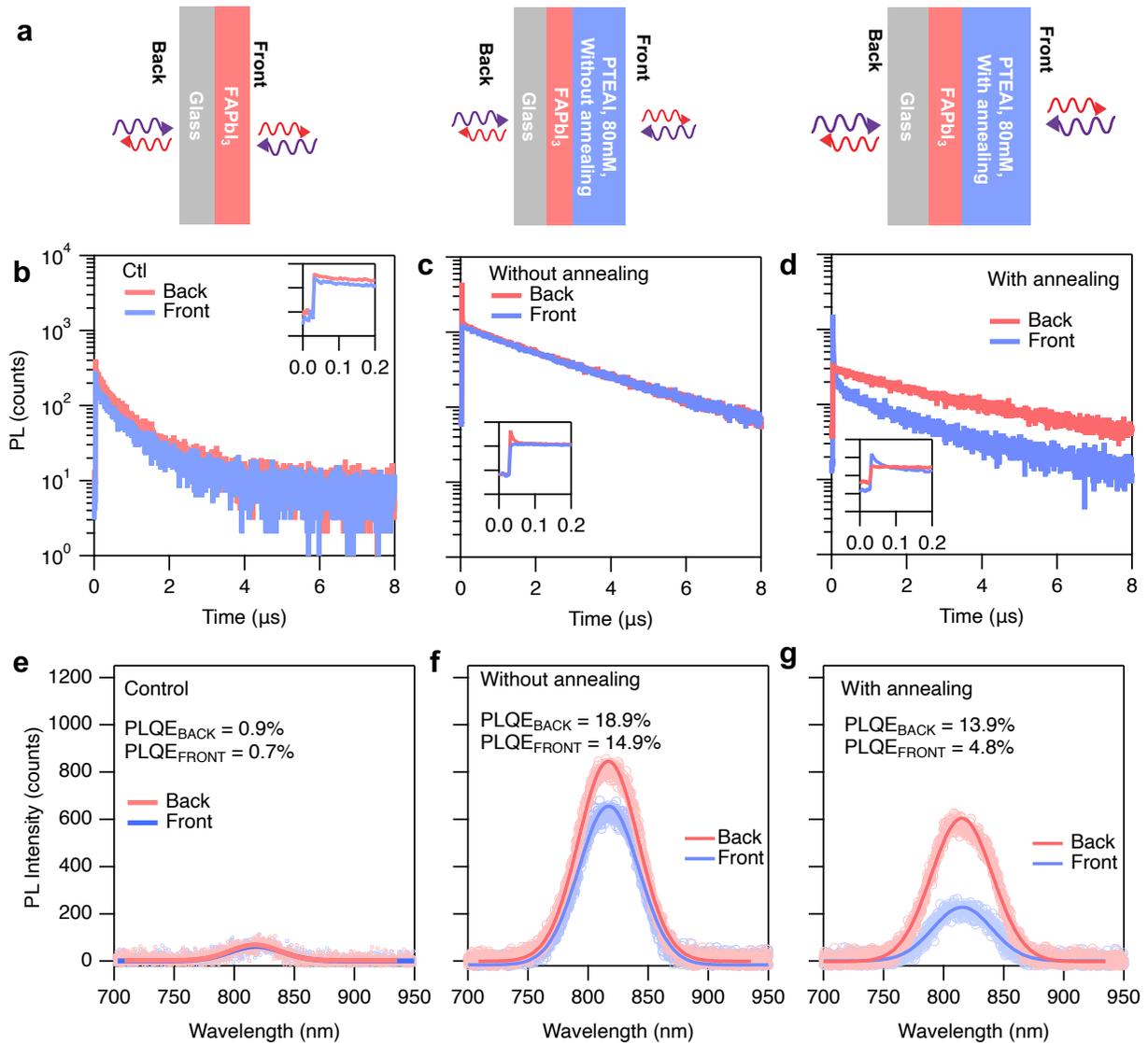

Fig. 1 Perovskite films and time-resolved photoluminescence (TRPL) and PL quantum efficiency (PLQE) measurements. Sample configuration of control (FAPbI$_3$), without annealing (PTEAI-80mM treated FAPbI$_3$), and with annealing (PTEAI-80mM treated FAPbI$_3$, 100°C for 10 min) (a) and resulting TRPL (b – control; c- without annealing; d – with annealing) and PLQE (e – control; f- without annealing; g– with annealing).

In the control samples (Fig. 1b), both the front excitation and back excitation show similar dynamics. In contrast, the PTEAI treated samples show different dynamics: without annealing (Fig. 1c), the films show a long carrier lifetime (2318 ns, Table S1) with no initial fast decay *via* front excitation whereas we observe a fast initial decay ($\tau_1$= 7.2 ns) and slow decay ($\tau_2$= 2169 ns) when excited from the back side. For the case with annealing (Fig. 1d), the PTEAI treated TRPL decay dynamics are the opposite, showing long carrier lifetimes (2786 ns) via back excitation and a fast initial decay ($\tau_1$= 14.6 ns) followed by a slow decay ($\tau_2$= 1428 ns) when excited from the front. For the samples without annealing, back excitation (through the glass) reveals a fast initial decay as compared to the front excitation, but the TRPL dynamics remains the same for the long lifetime component (Fig.1c). Interestingly, the fast decay component has a higher signal intensity than the initial value of front excitation, suggesting that is feature may originate through an extra radiative recombination channel.

External PL quantum efficiency (PLQE) measurements further confirm such an observation. The samples without annealing achieved PLQE values of 18.9% and 14.9% from back and front excitation, respectively. In comparison, the control sample PLQE values for the front and back are comparable ($PLQE_{back}$=0.9% and $PLQE_{front}$=0.7%). The samples with annealing demonstrated more dramatic differences between back and front excitation, with PLQE values from the back excitation much larger than front excitation ($PLQE_{back}$=13.9% and $PLQE_{front}$=4.8%), consistent with the TRPL decay dynamics. The observation of such a distinct carrier dynamic appears consistent across many AX/perovskite systems, including formamidinium-methylammonium-caesium triple cation perovskites (FAMACs) with PTEAI (Fig. S2), highlighting that these observations are general to a variety of perovskite systems.

Previously, the fast quench of TRPL from the AX treated side (i.e. "front side") is reported for some 2D/3D structure of perovskite,[27,39] but the fast quench from the back side with AX treatment has not previously been observed. This raises the question as to whether the fast initial drop in TRPL intensity observed for the annealed and not annealed samples originate from the same recombination mechanism (although observed from different sides of the sample). To answer this question, we performed streak camera measurements focused on the initial TRPL quench dynamics (Fig. S3). We found that in the $FAPbI_3$/PTEAI treated structures that were not annealed, the emission wavelength does not change with respective to the front or back excitation. But in samples with annealing, there is a long emission shoulder from the front excitation, which is absent from the back excitation, that disappears after ~1 ns, likely indicating a new emission species.

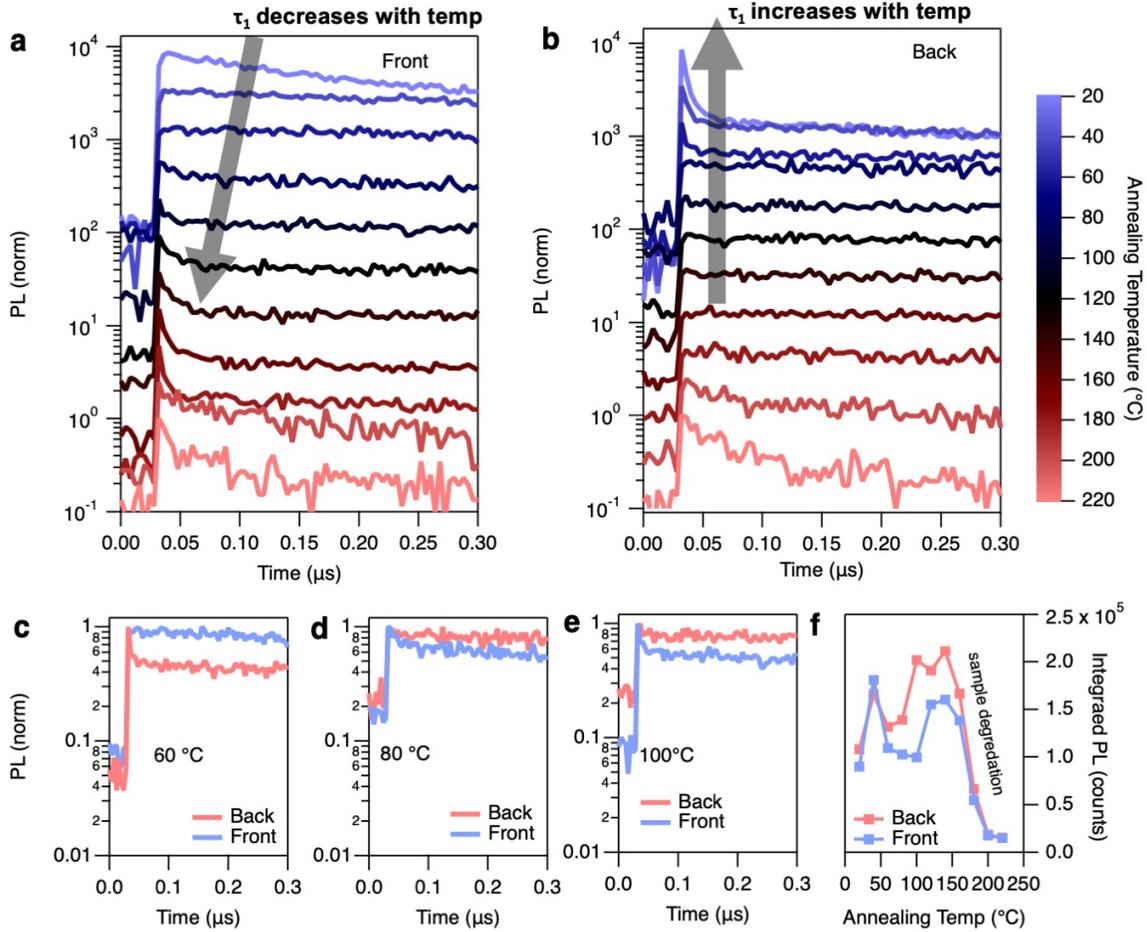

Fig. 2 Annealing temperature dependent TRPL. TRPL of PTEAI treated FAPbI$_3$ from the front excitation (a) and back excitation (b). The annealing temperature ranges from 40°C to 220°C. Representative TRPL traces at 60°C (c), 80°C (d), 100°C (e). (f) Integrated PL counts of the TRPL traces of samples that are annealed at different temperatures. The concentration of the PTEAI in the sample is 80 mM.

To better understand the excited state dynamics of FAPbI$_3$/PTEAI treated samples with and without annealing, we performed annealing temperature dependent TRPL measurements. In Fig. 2a and 2b, we show waterfall plots of the TRPL for front and back excitation for increasing annealing temperature. Consistent with Fig. 1, for non-annealed samples (i.e., at 20°C), we do not observe a fast-initial TRPL quench from the front excitation (Fig. 2a), while a fast quench is observed when excited from the back side (Fig. 2b). As the annealing temperature increases, we observe the fast initial quench from the back excitation (Fig. 2b) disappear while a fast quench appears from front excitation (Fig. 2a). The transition occurs, as shown in snapshots of three critical temperatures in Fig. 2c, d, e, between 60°C and 80°C. As the annealing temperature is increased, the fast quench from the front side becomes more pronounced (Fig. 2e). In Fig. 2f, we show the integrated PL counts from the TRPL traces as a function of annealing temperature. Interestingly, we observe similar functional behaviors between the two scenarios, where there are two regions of increasing and decreasing PL, with peaks at 40°C and 140°C. This indicates at least two distinct mechanisms of photoexcited dynamics present as the annealing temperature increases with PTEAI-treated FAPbI$_3$. While the decrease in PL after 140°C can be characterized by the degradation of the PTEAI-treated
8

FAPbI$_3$ interface as shown in PEAI-treated FAMACs,[40] the decrease in integrated PL intensity at 40°C warrants further investigation. As the transformation with temperature is likely related to thermally induced mass transport dynamics related to PTEAI, modulating the quantity of PTEAI on the FAPbI$_3$ should allow us to better understand the distinct TRPL dynamics. Therefore, we next evaluate the impact of PTEAI concentration on the PTEAI/FAPbI$_3$ system.

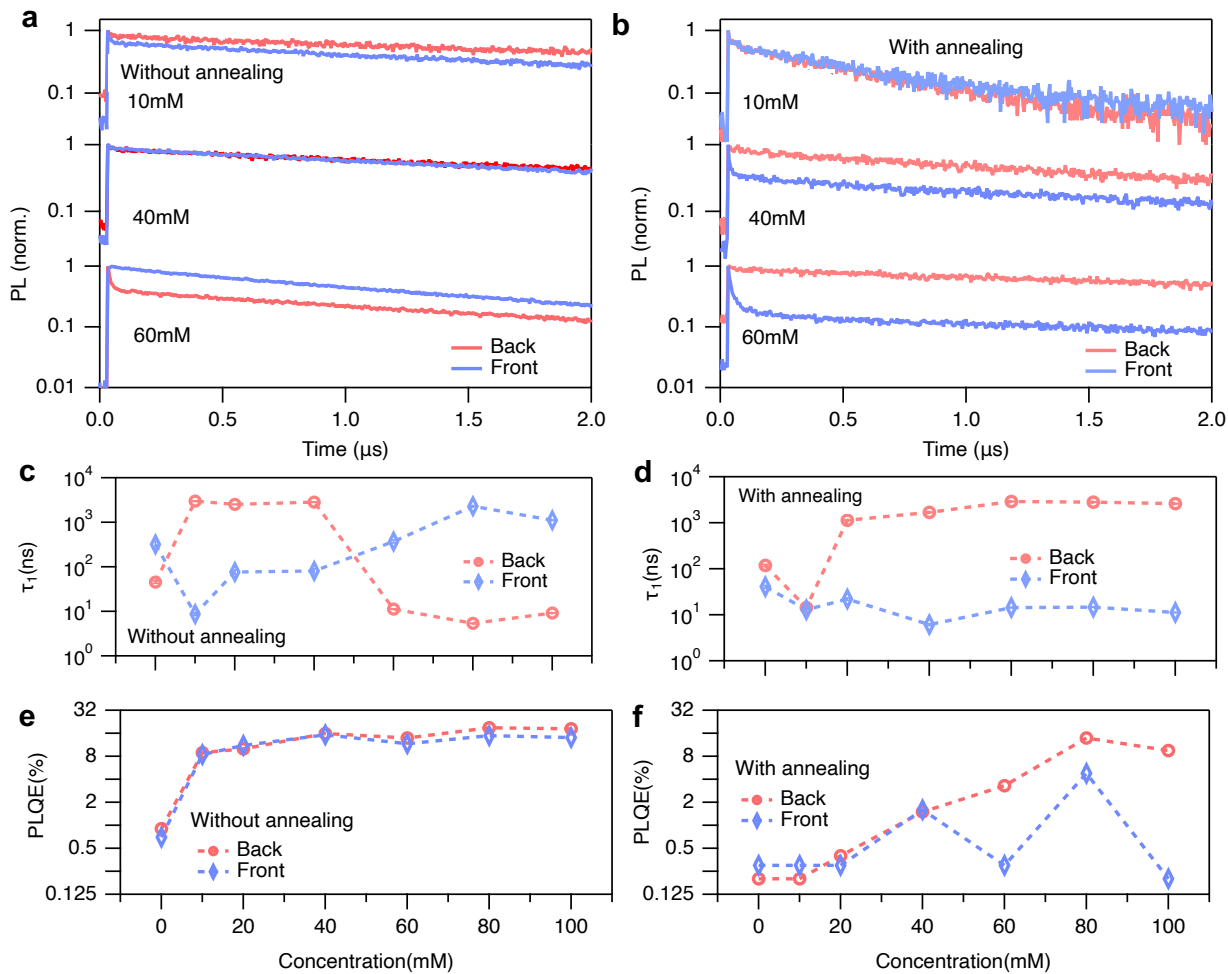

Fig. 3 PTEAI concentration dependent TRPL and PLQE without and with annealing (100°C, 10 min). TRPL at 10 mM, 40 mM, 60 mM PTEAI capping without annealing (a) and with annealing (b). Fast decay time fitting of the TRPL of PTEAI concentration ranging from 10 mM to 100 mM for samples without annealing (c) and with annealing (d). PLQE of PTEAI concentration ranging from 10 mM to 100 mM for the cases without annealing (e) and with annealing (f). Error bars are the standard deviations (S.D.).

Previously, optimal concentrations of AX have been reported anywhere from 1mM to 40 mM,[23,25–27,29–31] obtained through empirical device optimization. To understand the impact of PTEAI concentration on the photoexcited charge carrier dynamics, we tracked TRPL dynamics with PTEAI concentrations ranging from 10mM to 100mM (Fig. 3a, 3b, Fig. S4), and photoexciting from both the front and back surfaces. In addition to varying the PTEAI concentration, we also measured TRPL data for the scenarios without annealing and with annealing (i.e., 100 °C, 10 min). Surprisingly, without annealing, we observe a fast decay component with front excitation (Fig. 3a) that disappears as the concentration of PTEAI increases to and goes beyond 40mM (Fig. 3a, 3c). At the higher concentrations, a fast decay component from the back excitation is observed as we noted previously (Fig. 1c, Fig. 2). In the annealed samples, the fast initial decay of PL is observed from both front and back excitation at 10mM, but, at higher concentrations, the fast decay is only observed from the front side (Fig. 3b). To quantify these excitation direction dependent dynamics and PL intensities, the fast decay lifetime ($\tau_1$) and PLQE of the samples are presented in Fig. 3c, d, e, f and Table S2, Table S3. First, we found that, in the samples without annealing, the transition from observing no fast initial decay to having a fast initial decay occurs between 40 mM to 60 mM for backside excitation. With the emergence of the fast initial decay via back excitation, we do not see a corresponding dip in PLQE, but rather an increase in the PLQE as the fast initial decay becomes more prominent at higher concentrations. This suggests that the fast decay from the back excitation at higher treatment concentrations is due to enhanced radiative recombination. For the annealed samples, after 10 mM, the fast initial quench is always present for front excitation, while no such quench observed for back excitation.

To summarize the TRPL observations from above, we found that post-surface treatments of FAPbI$_3$ with PTEAI and no annealing step led to a fast initial quench from the back excitation while no such fast quench is observed from the front excitation. The nature of the initial quench appears to be radiative as the PL intensity and PLQE do not decrease with the advent of the fast initial quench. On the other hand, after 10 min at 100°C annealing post-treatment, a fast quench from the front side is observed regardless of the concentration of PTEAI, while back excitation does not show the fast component. These dynamics are highly dependent on the annealing temperature and treatment concentration. The temperature transition and concentration transition for observing such distinct TRPL dynamics is 60°C to ~80°C at 80mM and 40mM to 60mM at room temperature. To understand the origins of these process induced dynamics, we turn to X-ray diffraction (XRD) measurements for structural and compositional analysis.

**Structure and composition changes with PTEAI treatment of FAPbI$_3$**

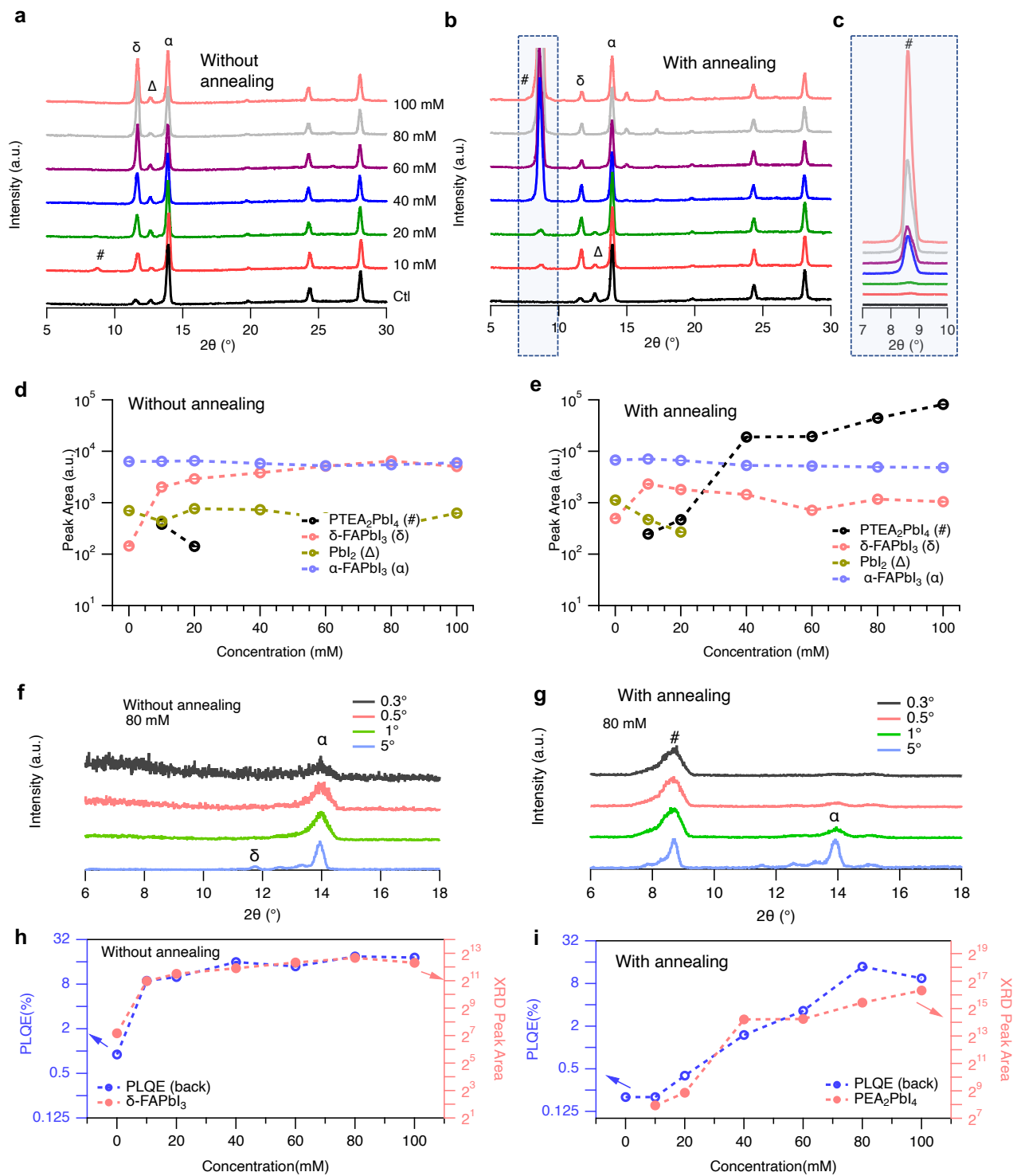

Fig. 4 PTEAI concentration dependent XRD without and with annealing (100°C, 10min). XRD of FAPbI$_3$/PTEAI where the concentration of PTEAI ranges from 10 mM to 100 mM with no further annealing (a) or with annealing (b,c). Key XRD peaks are identified as PTEA$_2$PbI$_4$ (labeled as "#"), PbI$_2$ (labeled as "Δ"), δ-FAPbI$_3$ (labeled as "δ"), α-FAPbI$_3$ (labeled as "α") and peak areas are tracked (d, e). Error bars are S.D.. GI-XRD of 80 mM PTEAI treated samples without annealing (f) and with annealing (g). PLQE and XRD peak area of δ-FAPbI$_3$ versus PTEAI concentration for samples that have no further annealing (h) and with annealing (i).

XRD was performed on samples without annealing and with annealing (10 min 100°C) after PTEAI treatment for concentrations ranging from 10 mM to 100 mM (Fig. 4a, b, c). Key XRD peaks are identified to be PTEA$_2$PbI$_4$ (8.6°)[34], PbI$_2$ (12.6°), δ-FAPbI$_3$ (11.6°) and α-FAPbI$_3$ (13.9°)[33], where we track their peak areas in Fig. 4d, 4e. Without annealing, surprisingly, as the PTEAI concentration increases, δ-FAPbI$_3$ content also has increased significantly, whereas α-FAPbI$_3$ content doesn't change, thus creating a δ/α-FAPbI$_3$ mixed state (Fig. 4d). Comparing the rise of δ-FAPbI$_3$ with correspondent film PLQE shows extraordinary similar trend (Fig. 4h), demonstrating that the PTEAI induced δ/α-FAPbI$_3$ mixed state is likely what leads to the PLQE enhancement in the samples that were not annealed. While δ-FAPbI$_3$ is mostly undesired[41–44] as photovoltaic material with its yellow color, the presence of δ/α-FAPbI$_3$ mixed state improves the film optoelectronic performance as observed by Qiao et al.[45] where δ/α-FAPbI$_3$ mixed state resulted in a 10-fold higher PLQE than the pure α-FAPbI$_3$ phase due to the formation of nanograins of δ/α-FAPbI$_3$ that enhances the radiative recombination.[46] Grazing incident -XRD (GIXRD) of an 80mM PTEAI treated FAPbI$_3$ without annealing show such a δ/α-FAPbI$_3$ mixed state is not formed at the top surface (Fig. 4f), which well explains the observation of fast radiative recombination at higher PTEAI concentration treatment when excited from the backside (Fig. 3c). Furthermore, for samples without annealing, PTEA$_2$PbI$_4$ only forms at low concentration treatment (< 20mM) (Fig. 4d), which aligns well with the observation of shorter τ$_1$ when excited from the front side (Fig. 3c). For samples with higher concentrations PTEAI (> 20 mM) treatment and without annealing, we do not observe any new crystalline species, consistent with amorphous PTEAI surface layers (Fig. S5).

In comparison, with annealing, the PTEA$_2$PbI$_4$ peak area increases dramatically as the PTEAI concentration increases (Fig. 4b, 4c, 4e). At the same time, PbI$_2$ disappears (Fig. 4e) as more PTEA$_2$PbI$_4$ forms with annealing, suggesting the PbI$_2$ is the main source for forming PTEA$_2$PbI$_4$, as compared to a cation exchange through α-FAPbI$_3$ (Fig. 4e) or δ-FAPbI$_3$ as suggested by Liu et al.[47] GIXRD of an 80mM PTEAI treated FAPbI$_3$ shows the PTEA$_2$PbI$_4$ is mostly concentrated at the surface (Fig. 4g). The appearance of PTEA$_2$PbI$_4$ correlates well with the fast decay of TRPL from the front excitation (Fig. 1, Fig. 3), indicating that the fast quench in TRPL from the front excitation is a ubiquitous photoexcited dynamic for the FAPbI$_3$/PTEA$_2$PbI$_4$ system, with possible extensions to other 3D/2D halide perovskite systems. Indeed, the front excitation quench in TRPL with 3D/2D is reported by many reports such as Nazeeruddin et al[39] and Sargent et al[27], and there are several potential mechanisms that could explain this fast component including charge and energy transfer dynamics across the 2D/3D interface. While the exact nature of such a transfer at the 2D/3D perovskite require further research, the possibility of using TRPL to indirectly detect the presence of 2D perovskite on the 3D perovskite surface is a highly useful method. Indeed, this approach could be applied to rapidly detect 2D/3D perovskite interfaces on 3D/2D/glass structures, which are traditionally hard to detect using other characterization methods.

**Morphological and spatial variations with PTEAI treated FAPbI₃**

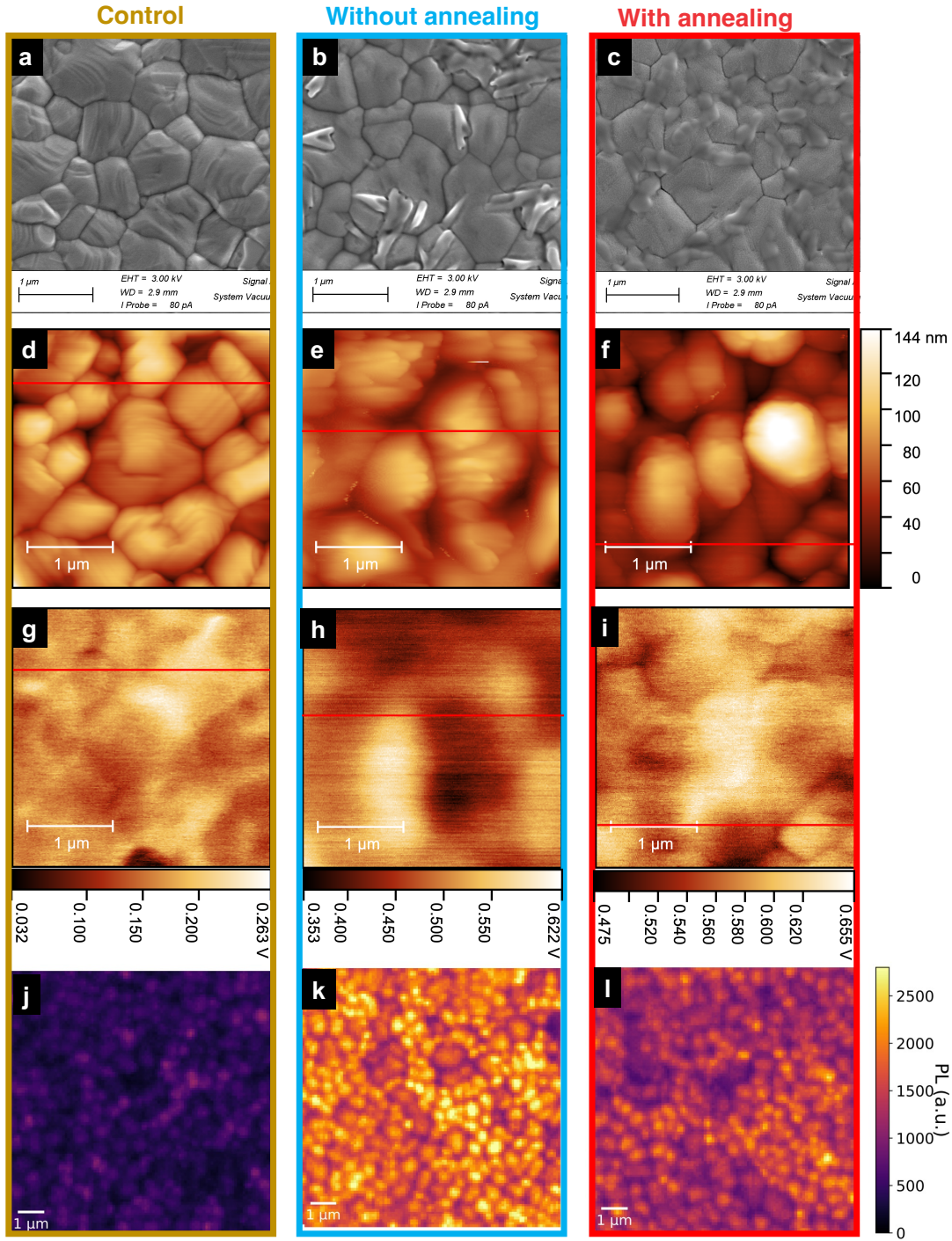

Fig. 5 Morphology and correlated field potential, photoluminescence of PTEAI (40mM) treated FAPbI$_3$. Surface SEM of FAPbI$_3$ film (a), PTEAI treated FAPbI$_3$ without annealing (b), PTEAI treated FAPbI$_3$ with annealing (c); AFM of FAPbI$_3$ film (d), PTEAI treated FAPbI$_3$ without annealing (e), PTEAI treated FAPbI$_3$ with annealing (f); KPFM of FAPbI$_3$ film (g), PTEAI treated FAPbI$_3$ without annealing (h), PTEAI treated

FAPbI$_3$ with annealing (i); PL map of FAPbI$_3$ film (j), PTEAI treated FAPbI$_3$ without annealing (k), PTEAI treated FAPbI$_3$ with annealing (l).

To further understand the impact of the PTEAI modifications on the surface morphology of the films and their implications on the film performance, we performed scanning electronic microscopy (SEM) and atomic force microscopy (AFM) measurements, and the results are compared to Kelvin probe force microscopy (KPFM) and confocal PL images. We paid particular attention to the grain boundaries where PbI$_2$ most likely resides.[48] Figure 5a, b, and c show top-view SEM images of FAPbI$_3$, PTEAI/FAPbI$_3$/glass without and with annealing. In Fig. 5b we observe small features with larger contrast, presumably due to the insulating character of the domains indicative of charging effects. Since they were not present in Fig.5a, we believe these are PTEAI related domains or PbI$_2$. After annealing (Fig. 5c), plate-like features appear that can be attributed to the formation of PTEA$_2$PbI$_4$ through PbI$_2$ + 2PTEAI ⇋ PTEA$_2$PbI$_4$ (Fig. 5c).

Correlating the surface topography of AFM (Fig. 5d-f) with the contact potential difference (CPD) of KPFM (Fig. 5g-i, Fig. S6) show an anticorrelation, particularly at the grain boundary, for control samples. Such an observation is consistent with KPFM measurements with other perovskite materials systems (FACs, FAMACs)[49]. In contrast, for the annealed samples, the grain boundary CPD has changed to become positively correlated with the grain boundary locations, further validating that PTEAI annealing modifies the top surface and grain boundaries. For non-annealed samples, the surface topography correlation with CPD does not change demonstrably from the control samples (Fig. 6b), implying the film top surface morphology and electrical potential is mostly preserved with the treatment. Finally, the impact of the modifications on the film performance is assessed with confocal PL mapping (excited from the back side). It shows that, with the PTEAI treatment (without annealing and annealing), the film's photoluminescence increases significantly (Fig. 5j-i). Samples without annealing show much higher PL than samples that have been annealed, which is consistent with our previous observation showing much higher PLQE without annealing *via* back excitation, despite having a fast initial lifetime component.

## Proposed mechanism of PTEAI treated FAPbI$_3$

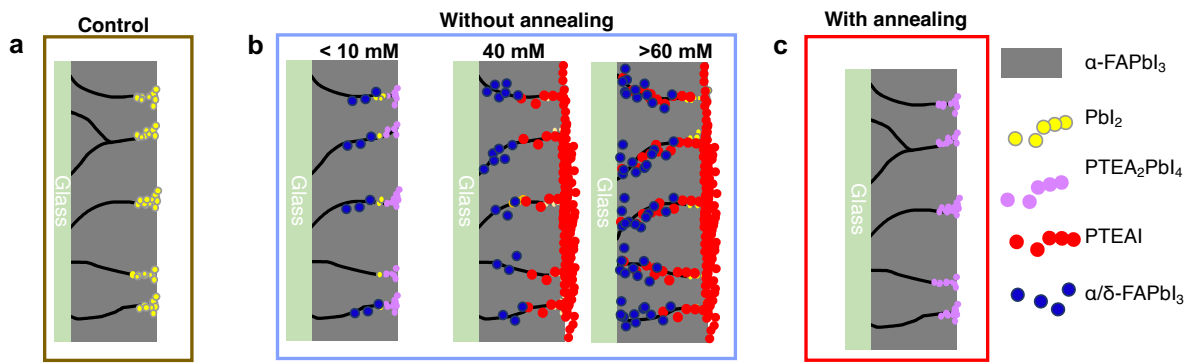

Fig. 6 Proposed mechanism to describe the unique excited dynamics of PTEAI treated FAPbI$_3$ systems. FAPbI$_3$ (a), PTEAI treated FAPbI$_3$ without annealing (b) and PTEAI treated FAPbI$_3$ with annealing (c).

Combining our understanding of excited carrier dynamics, structural dynamics, and morphology for the FAPbI$_3$/PTEAI system, we propose a mechanism in Fig. 6 that summarize these results. The control FAPbI$_3$ films contains surface defects, including at the grain boundaries where PbI$_2$ present, and which act as trap centers for nonradiative recombination (Fig. 6a). Upon treating the FAPbI$_3$ surface with PTEAI at low concentrations (<20mM), PbI$_2$ at the surface reacts with PTEAI to form PTEA$_2$PbI$_4$. As a result, some of the surface traps are passivated, and a long carrier lifetime is observed when exciting from the back side. For front excitation, the presence of PTEA$_2$PbI$_4$ quenches the PL signal. As the concentration of PTEAI increases, instead of forming PTEA$_2$PbI$_4$, PTEAI induces the formation of δ/α-FAPbI$_3$ phase in the bulk and bottom side that allows a fast radiative recombination and enhance the photoluminescence quantum yield. For samples that are annealed at 100°C for 10 min, PTEAI is converted to PTEA$_2$PbI$_4$ regardless of the PTEAI concentration. PbI$_2$, which is localized largely at the top surface and near-surface grain boundaries, is the source of lead for PTEA$_2$PbI$_4$, and thus the location and quantity of PbI$_2$ determines where PTEA$_2$PbI$_4$ forms. Therefore, in the case where the sample is annealed, PTEA$_2$PbI$_4$ is formed mostly at the top surface of FAPbI$_3$.

## Impact of PTEAI treatments on device $V_{OC}$

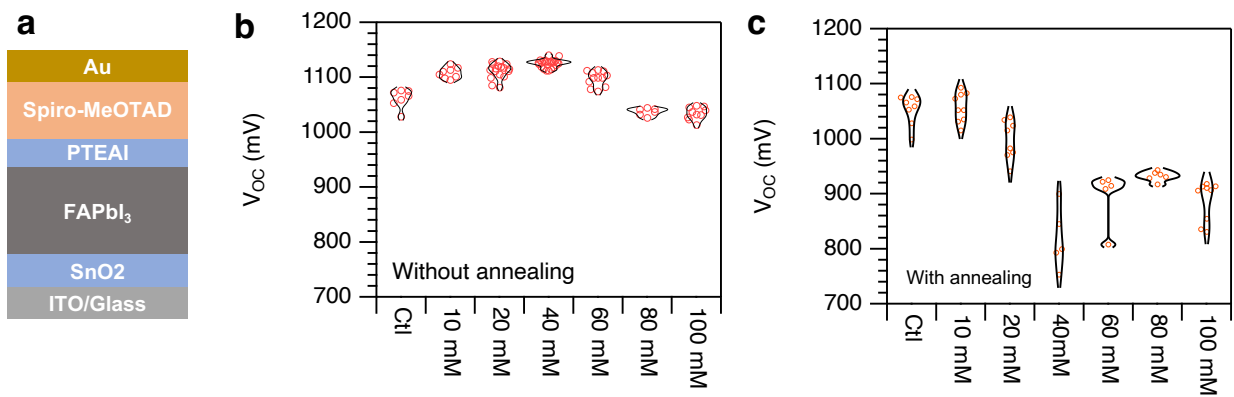

Fig. 7 Device architecture (a) and resulting device open-circuit voltages without annealing (b) and with annealing (100°C for 10 min) (c). The shape of plot represents the population size of the data.

To understand the implications of the interfacial charge dynamics highlighted in this study, we fabricated solar cells with the following structure: Glass/ITO/SnO$_2$/FAPbI$_3$/PTEAI/Spiro-OMeTAD/Au (Fig. 7a). Through reducing nonradiative recombination, AX passivation of the perovskite mostly impacts the open-circuit voltage (V$_{OC}$) of devices[23,24,26,50]. Therefore, we primarily focus on the impact of these treatment on device V$_{OC}$. In addition, PL is connected to V$_{OC}$ through the reciprocity relations,[51–53] and therefore serves as an interrelated parameter to connect to our previous observations. The resulting device statistics from the current density – voltage characteristics are presented in Fig. S7-S8, Table S1-S2. While both the annealed and non- annealed samples show improvements in V$_{OC}$, the non- annealed devices show a much wider range of treatment concentrations and more reproducible V$_{OC}$ improvement (10-60mM), with the maximum improvement shown at 40mM (Fig. 8b). In contrast, with annealing, the device V$_{OC}$ improves only moderately at 10mM and decreases significantly with increasing concentration (Fig. 8b). Without annealing, PTEAI is broadly distributed at the surface and grain boundaries of the film, and thus resulting in greater opportunities for trap passivation, more uniquely, allowing δ/α-FAPbI$_3$ phase mixing of perovskite at perovskite/ETL interface. However, in the annealed samples, as PbI$_2$ determines the place and amount of PTEAI reacting to form PTEA$_2$PbI$_4$, the ability of PTEAI to passivate traps is largely limited by the surface and surface grain boundary compositional distributions.

In conclusion, we have identified two distinct interfacial dynamics in AX surface treated perovskite, taking PTEAI treated FAPbI$_3$ as a model system, with respect to the presence of AX (e.g., PTEAI) or layered perovskite (e.g., PTEA$_2$PbI$_4$) in the film. While both PTEAI and PTEA$_2$PbI$_4$ passivate traps in perovskites, the impact on the surface and interface of glass/FAPbI$_3$ is quite different. PTEAI can be present across the surface/interface and grain boundaries, acting as passivation agent and induce δ/α-FAPbI$_3$ phase mixing near the interface of glass/FAPbI$_3$. In comparison, PTEA$_2$PbI$_4$ is mostly formed at the surface and grain boundaries near the top surface. These differences have significant impact on the device performance where PTEAI shows a wider treatment concentration window and more reproducible V$_{OC}$ improvements, thus increases PCE. In comparison, formation of the low dimensional perovskite PTEA$_2$PbI$_4$ limits the range of concentrations where V$_{OC}$ enhancements are observed. By directly correlating the structural dynamics and carrier dynamics, we have provided a fast and simple technique to determine the type of interfacial layer formed and how changes in charge carrier dynamics translate to device performance. These results have major implications for the improvement of device performance and provide new TRPL metrics for diagnostic thin film photophysical characterization before and during device optimization.

**Author contributions**
B.D.D. conceived the idea and designed the experiments. B.D.D., D.W.D., M.L., R.B. interpreted experiment results. B.D., M.L., R.B., L.W., E.L.W., R.S., M.S. performed material and device characterizations. J.J.Y. N. T. P. H., S. W. contributed to the discussion. T.B., M.G.B., and V.B. supervised the study. B.D.D. wrote the first draft which was edited by D.W.D. and M.L., and all authors contributed feedback.

**Conflicts of interest**
V.B. is an advisor to Swift Solar, a US company developing perovskite photovoltaics and the co-founder of Ubiquitous Energy, a US company developing transparent photovoltaics. D.W.D. is a co-founder of Optigon, a US company developing metrology tools for the photovoltaics industry. All others have no conflicts to declare.

**Acknowledgements**

This is work is supported by MIT-Tata GridEdge Solar Research Program. This material is based upon work supported by the U.S. Department of Energy, Office of Energy Efficiency and Renewable Energy (EERE), under Award Number DE-EE0009512. M.L., E.W., and R.B. acknowledge support from the National Science Foundation Graduate Research Fellowship under Grant No. 24 (1122374).

Supplementary Information

**Processing Induced Distinct Charge Carrier Dynamics of Bulky Organic Halide Treated Perovskites**

Benjia Dak Dou [a, b], Dane W. deQuilettes [a, b], Madeleine Laitz [a, b], Roberto Brenes [a,b], Lili Wang [c], Ella L Wassweiler [a,b], Richard Swartwout [a,b], Jason J. Yoo [c, d], Melany Sponseller [a,b], Noor Titan Putri Hartono [e], Shijing Sun [e], Tonio Buonassisi [e], Moungi G Bawendi [c], Vladimir Bulović [a,b,*]

[a] Research Laboratory of Electronics, Massachusetts Institute of Technology, 77 Massachusetts Avenue, Cambridge, Massachusetts 02139, USA. bulovic@mit.edu
[b] Department of Electrical Engineering and Computer Science, Massachusetts Institute of Technology, 77 Massachusetts Avenue, Cambridge, Massachusetts 02139, USA.
[c] Department of Chemistry, Massachusetts Institute of Technology, 77 Massachusetts Avenue, Cambridge, Massachusetts 02139, USA.
[d] Division of Advanced Materials, Korea Research Institute of Chemical Technology, Daejeon, Republic of Korea
[e] Department of Mechanical Engineering, Massachusetts Institute of Technology, 77 Massachusetts Avenue, Cambridge, Massachusetts 02139, USA

EXPERIMENTAL METHODS

**Materials**

Unless otherwise noticed, all the materials and solvents are from Sigma-Aldrich. Formamidinium Iodide (FAI), Methylammonium Chloride (MACl), and are from GreatCell Solar. Lead(II) Iodide (PbI2, 99.99%) is from Tokyo Chemical Industry (TCI). SnO2 nanoparticle ($SnO_2$) is from Alfa Aeser. Spiro-MeOTAD (LT-S922) is from Lumtec. Prepatterned indium-doped Tin Oxide (ITO) substrates are from Thin Film Devices.

**Film preparation**

$FAPbI_3$ films were fabricated from $FAPbI_3$ powders following the Park et al.[1] Briefly, PbI2 and FAI ($PbI_2$:FAI = 1:1.3) is mixed in acetonitrile with a concentration of 0.1g/mL. The resulting solution is stirred for 48 hours at room temperature and then washed 3 times with acetonitrile and recovered. To prepare $FAPbI_3$ solution for the film, 1000 mg $FAPbI_3$ powder and 20mg MACl are dissolved in 104 μL DMSO and 831 μL DMF, vortex for 2 min, and heated on a hotplate at 70°C for 60 min. A PTFE filter with 0.45 μm was used to filter the solution before making the films. To make $FAPbI_3$ films, the solution is spin-coated on the substrate at 1000 rpm for 10 s and 5000 rpm 30 s; 650 mL diethyl ether was drop at 20 s before the end of the spinning. The resulting films are annealed at 150°C for 10 min on hotplate. All of the solution preparation and film preparation were performed in a dry air glovebox of <5% relative humidity. For PTEAI treated films, 10mM - 100mM PTEAI solution in IPA is spin coated on the perovskite at 5000 rmp for 30 s with the acceleration of 5000rmp/s. The films are not further annealed or annealed with conditions discussed in the main texts.

**Device preparation**

ITO substrates are cleaned as follows: ultrasonic bath in acetone and isopropanol for about 10 min respectively; UV-ozone treatment for 20min. $SnO_2$ nanoparticle solution is diluted with DI water to 3 % (wt/wt) with 3 mg/mL KCl as the additive. The resulting solution is spin coat at 3000 rpm for 30 s with acceleration of 3000rpm/s and annealed at 150°C for 30 min. Perovskite films are prepared as in the previous session. Spiro-MeOTAD solution contains 72 mg Spiro-MeOTAD, 1 mL Chlorobenzene, 17.5 μL Li-TFSI stock solution (520 mg/mL Li-TFSI in acetonitrile) and 28.8 μL t-BP. The solution is then spin-coated at 4000 rpm (4000 rpm/s acceleration) for 30 s on the perovskite. Finally, a 70 nm for Au is thermally evaporated on the Spiro-MeOTAD to finish the device.

**Film Characterization**

Time resolved photoluminescence traces were collected using a homebuilt system where a 405 nm (3.06 eV) picosecond pulsed diode laser (Picoquant; LDH-DC-405) is applied to excite the films. The repetition rate is 125 kHz, laser spot size is 10.7 μm$^2$, and energy density (E $_\lambda$) is 1.86 nJ/cm$^2$/pulse. The resulting PL traces are fitted with a biexponential equation of $I(t) = A_1 \exp(-t/\tau_1) + x\, A_2 \exp(-t/\tau_2)$, where $\tau_1$ and $\tau_2$ are the fast and slow lifetime components of the decay, and $A_1$ and $A_2$ are the weighted ratios of $\tau_1$ and $\tau_2$, respectively. To avoid overfitting, the fitting is limited to single exponential if the biexponential fitting gives a large fitting error due to the lack of the fast lifetime.

PLQE measurements were done with well-established method from de Mello *et al*[2] where the film are excited with a 405 nm at 100mW cm$^{-2}$. PL map were collected using a custom-built TCSPC confocal microscope setup with a 100x oil objective as described in Brenes *et al* [3]. The films were excited using a 405 nm at 100mW cm$^{-2}$.

XRD and GIXRD were taking using a Rigaku SmartLab. SEM were taking using a ZEISS GeminiSEM. KPFM measurements were performed using an Asylum Research Cypher system.

**Device Characterization**
*J-V* characteristics of the devices were measured in ambient environment with a Keithley 2400 source meter under AM1.5 100 mW·cm$^{-2}$ irradiance. The light source is calibrated with encapsulated Si cell certified by the National Renewable Energy Laboratories (NREL). The device areas are 0.15 cm$^2$, defined by a metal mask. The J-V curves were measured from 1.3 V to -0.2 V and then backwards.

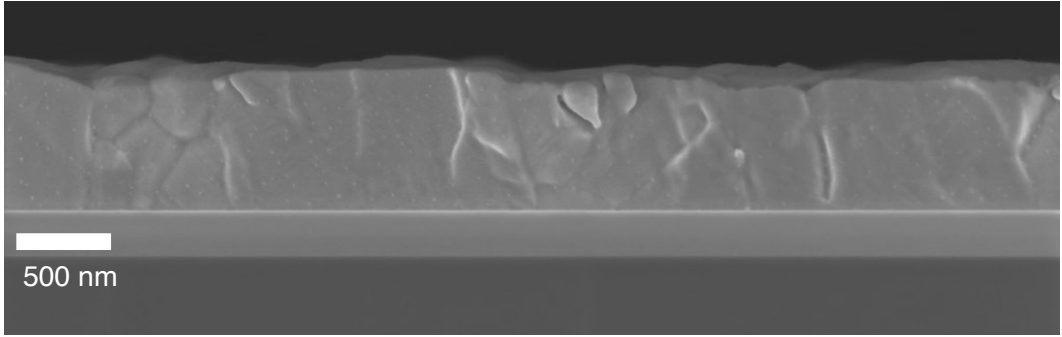

Fig. S1 FAPbI$_3$ films on Si wafer. The thickness of FAPbI$_3$ is *ca*. 740 nm.

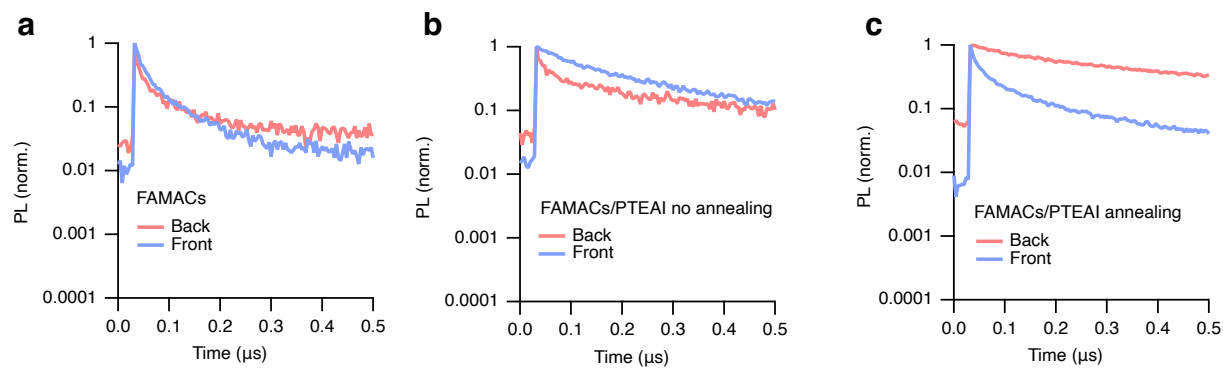

Fig. S2 TRPL of PTEAI treated FAMACs system. Control (a), PTEAI treated FAMACs wihtout annealing (b), and PTEAI treated FAMACs with annealing (c).

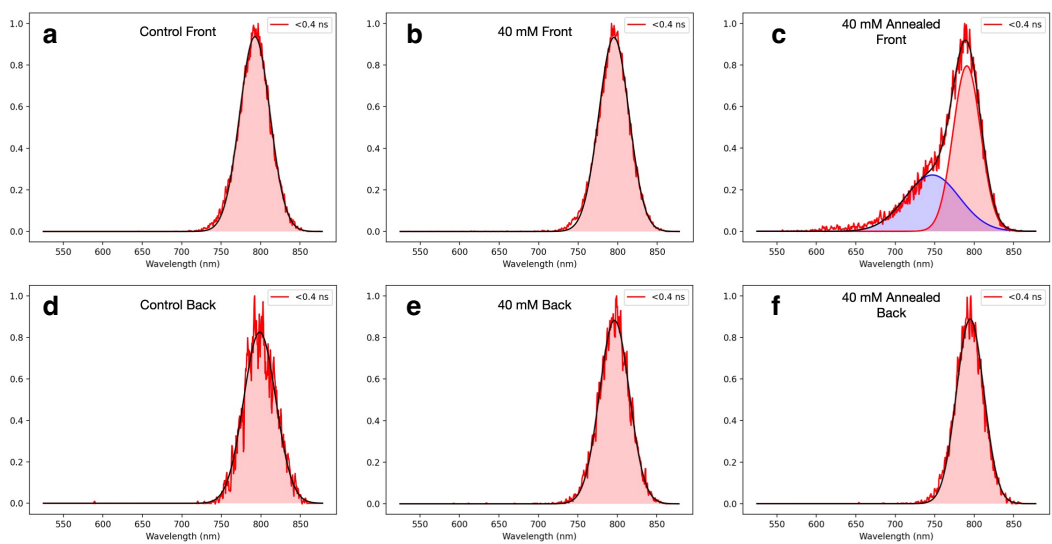

Fig. S3 Snapshot of the streak camera measurement of FAPbI$_3$/PTEAI. (a) Control, Front excitation; (b) PTEAI treatment without annealing, Front Excitation; (c) PTEAI treatment with annealing, Front Excitation; (d) Control, Back Excitation; (b) PTEAI treatment with annealing, Back Excitation; (c) PTEAI treatment with annealing, Back Excitation.

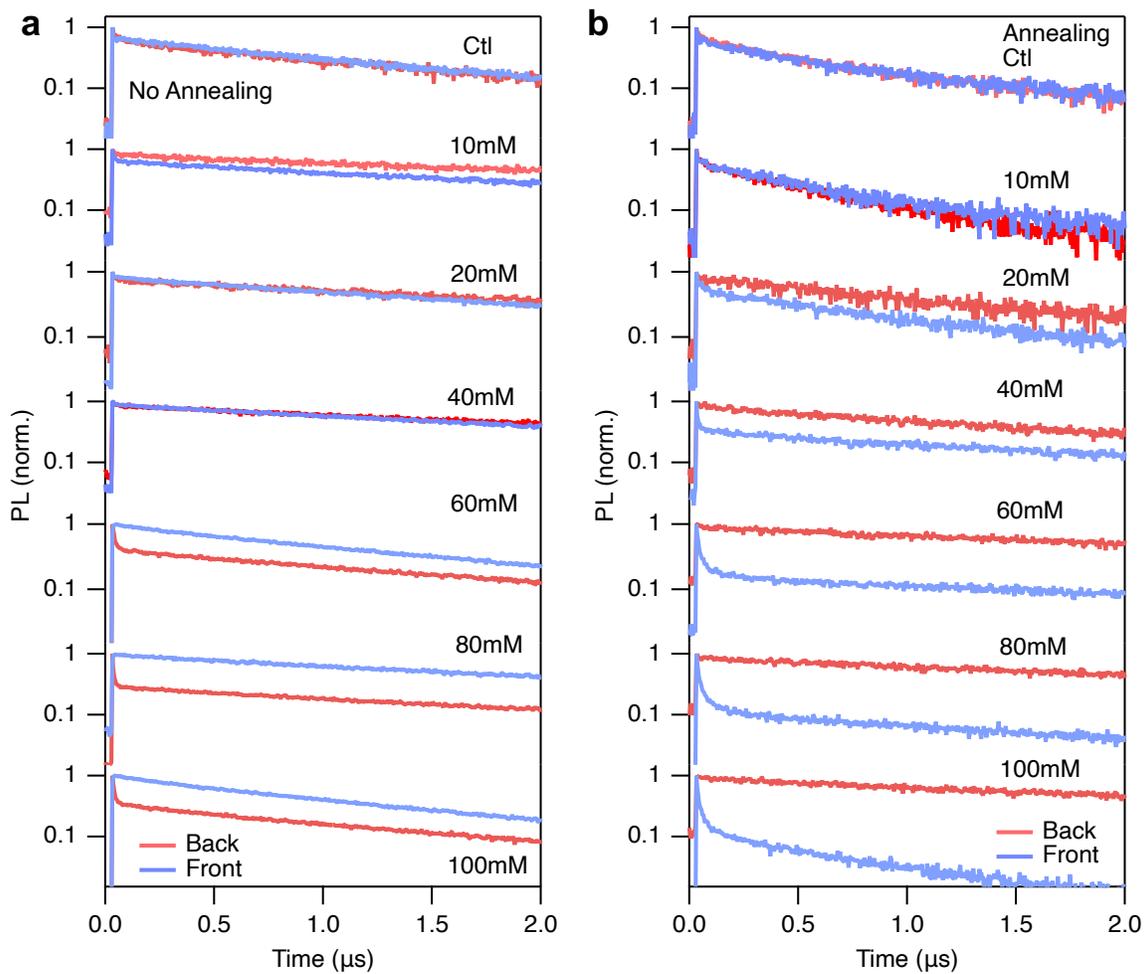

Fig. S4 TRPL traces of PTEAI treated FAMACs with the variation of PTEAI concentration (10mM – 100mM), and without (a) and with (b) annealing afterwards.

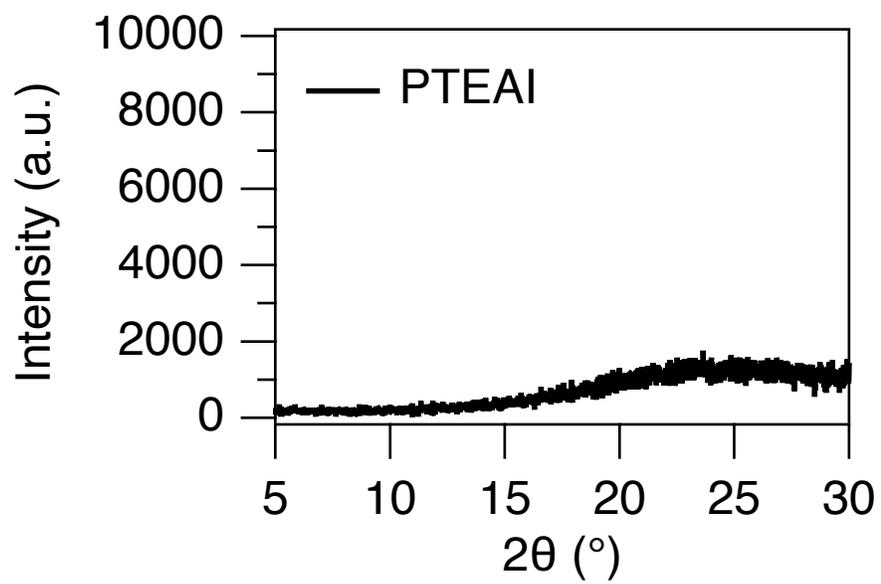

Fig. S5 XRD of PTEAI on glass.

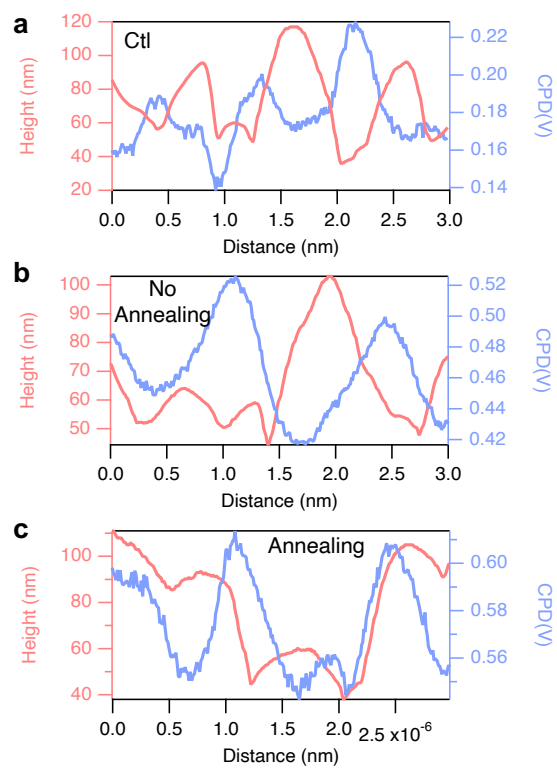

Fig. S6 KPFM grain boundary profiling. Control (a), PTEAI treated FAPbI$_3$ without annealing (b), and PTEAI treated FAPbI$_3$ with annealing (c).

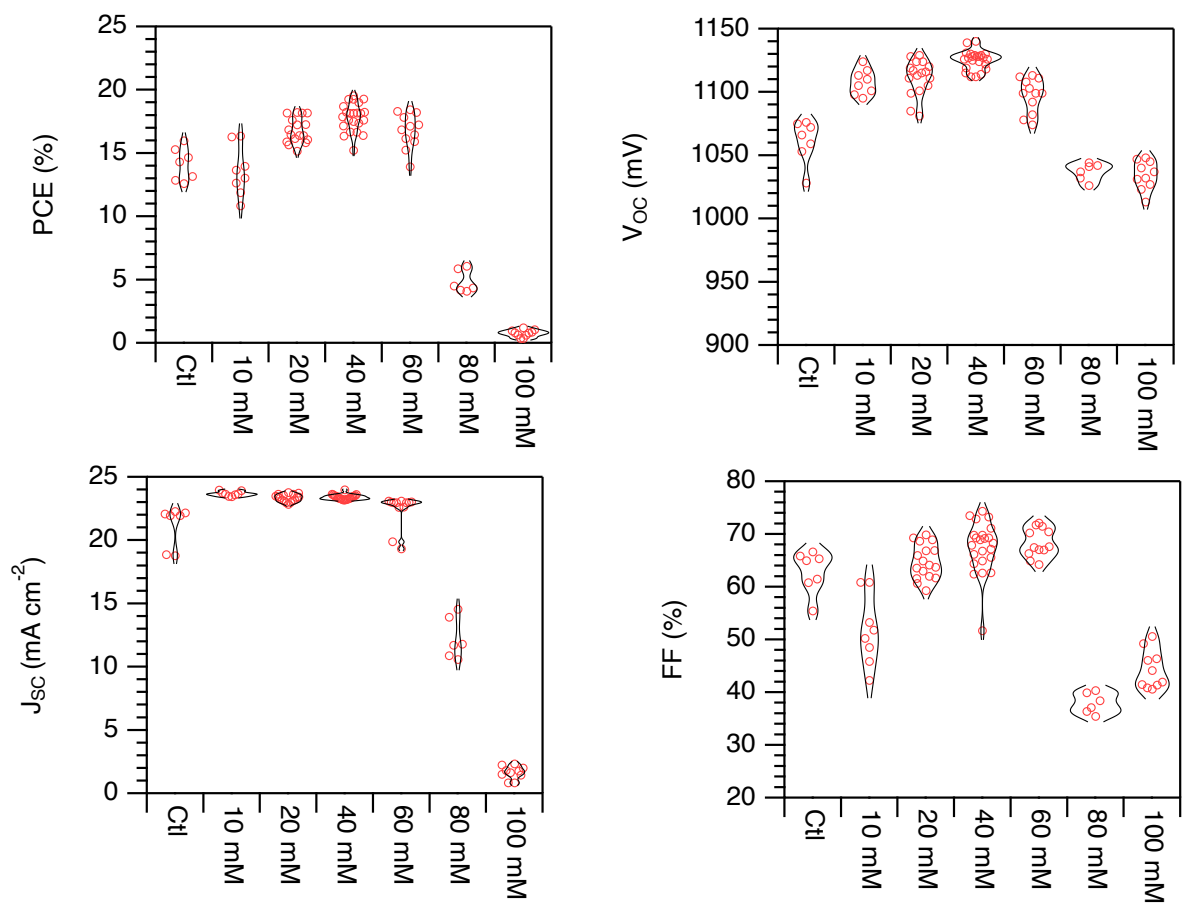

Fig. S7 PTEAI treated FAPbI$_3$ without annealing device. PCE (a), V$_{OC}$ (b), J$_{SC}$ (c), FF(d). The shape of plot represents the population size of the data.

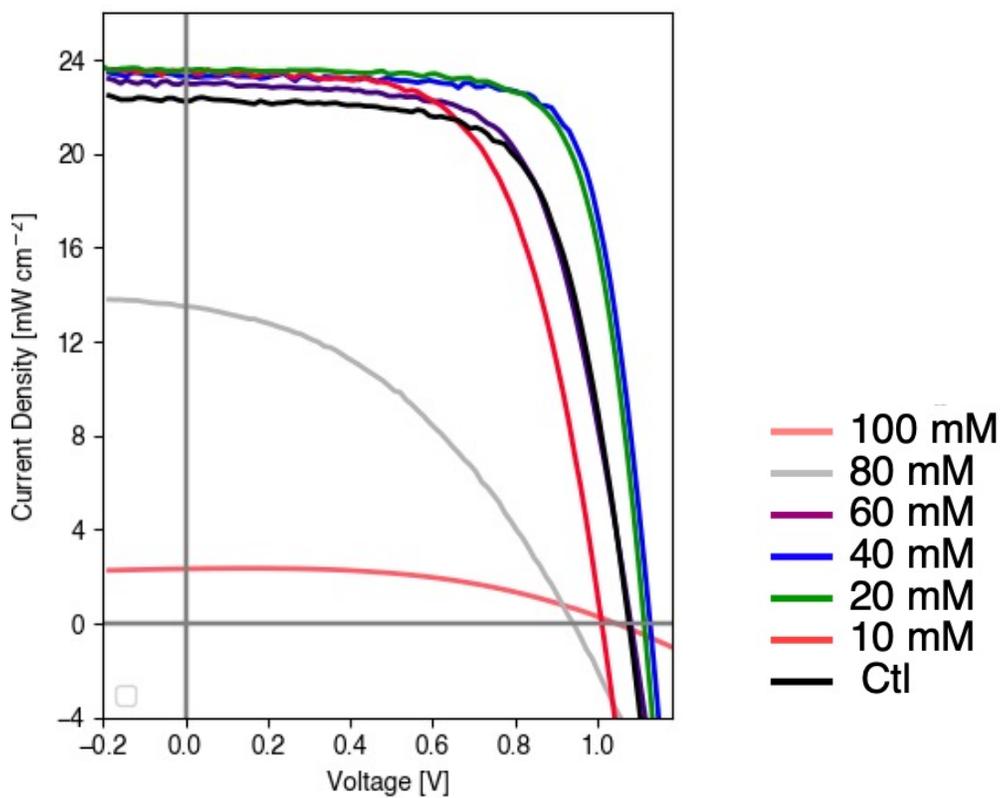

Fig. S8 *J-V* curves of the champion devices with variation of PTEAI concentration (10mM – 100mM), and with no annealing after the PTEAI treatment.

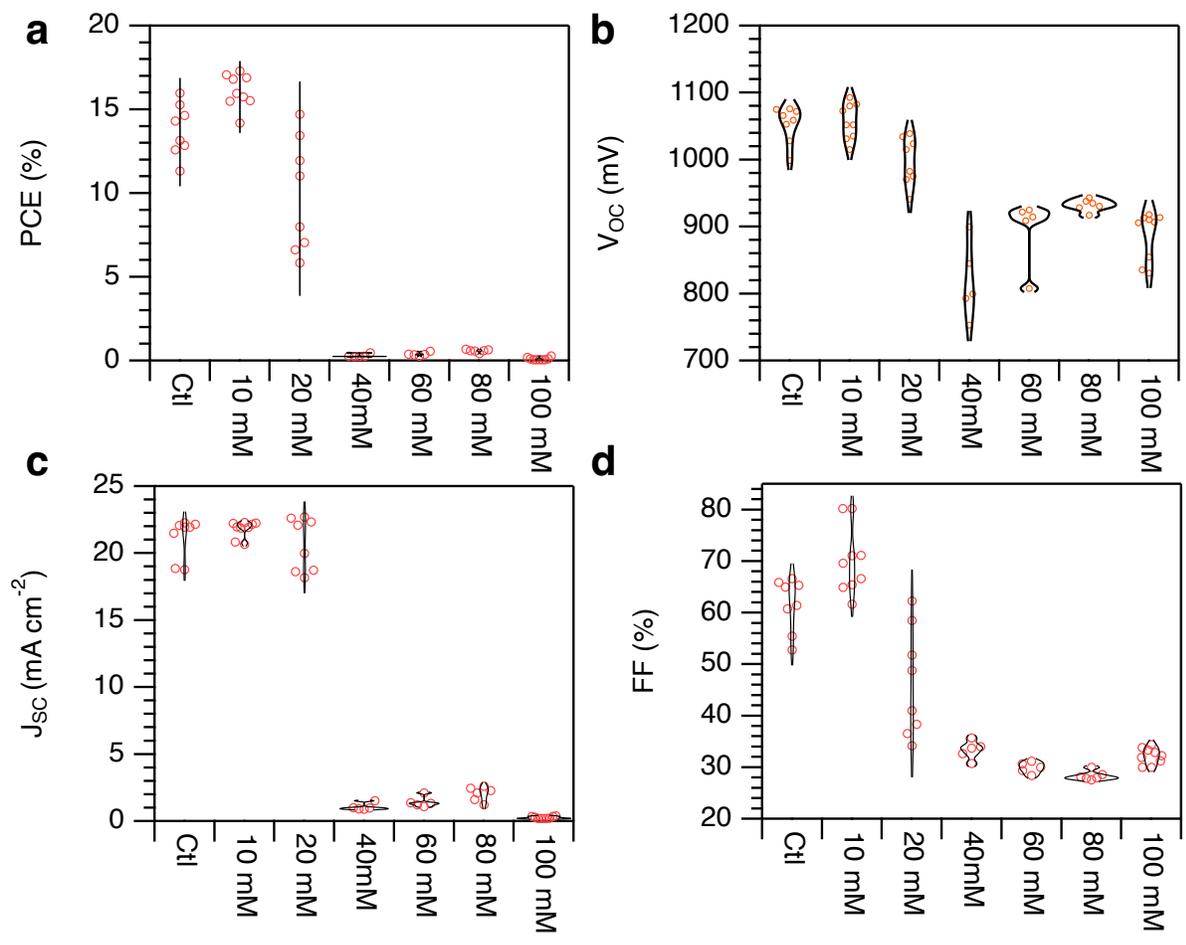

Fig. S9 PTEAI treated FAPbI$_3$ with annealing device. P: PCE (a), $V_{OC}$ (b), $J_{SC}$ (c), FF(d). The shape of plot represents the population size of the data.

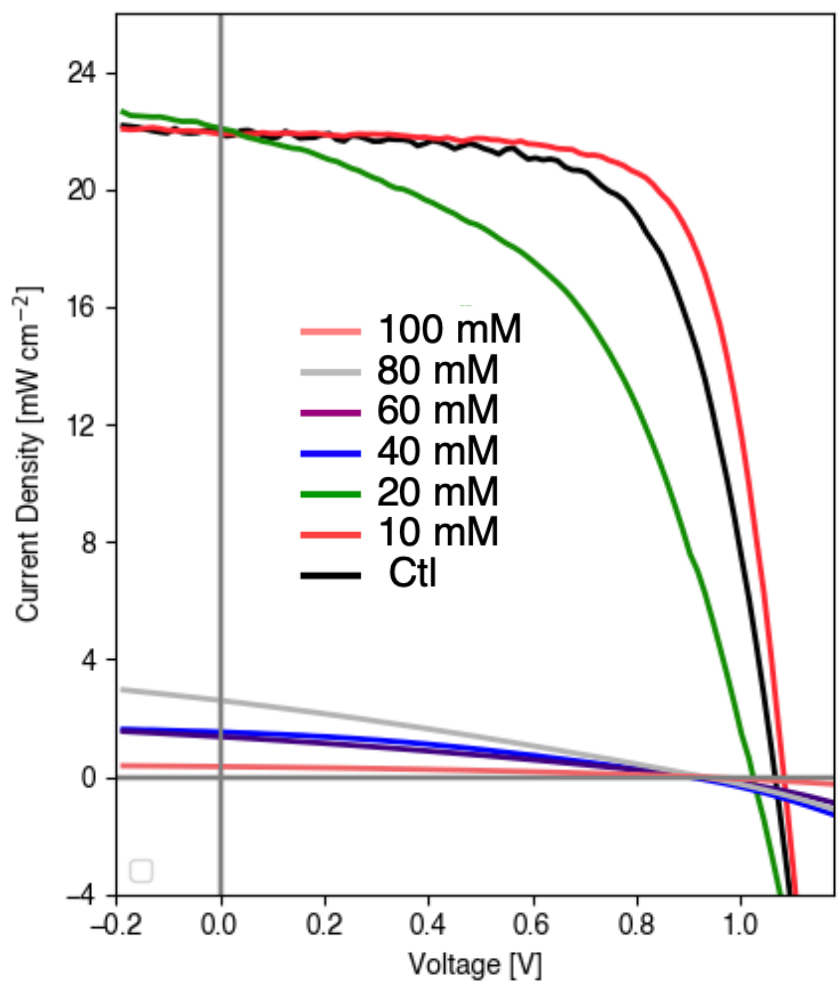

Fig. S10 *J-V* curves of the champion devices with variation of PTEAI concentration (10mM – 100mM), and with annealing after the PTEAI treatment.

Table S1 Biexponential fitting of the TRPL Traces in Fig. 1b, c, d

| | | $A_1$ (%) | $\tau_1$ (ns) | $A_2$(%) | $\tau_2$ (ns) |
|---|---|---|---|---|---|
| Control | Front | 37.2 | 40±2 | 62.8 | 784 ± 9 |
| | Back | 37.4 | 118 ± 6 | 62.6 | 810 ± 11 |
| No Annealing | Front | 2318 +11 | | | |
| | Back | 52.8 | 7.2 ± 0. 2 | 47.2 | 2169 ± 12 |
| Annealing | Front | 88.2 | 14.6 ±0.2 | 11.8 | 1428 ± 19 |
| | Back | 0 | 0 | 100 | 2786 ± 37 |

Table S2 Biexponential fitting of the TRPL Traces (without annealing) in Fig. S4a and Fig. 3a.

| Concentration | | $A_1$ | $\tau_1$ | $A_2$ | $\tau_2$ |
|---|---|---|---|---|---|
| Ctl | Front | 15.0 | 320.1±50.3 | 85.0 | 1367.5±25.6 |
| | Back | 33.6 | 45.4±2.9 | 66.3 | 1320.6±11.4 |
| 10 mM | Front | 37.2 | 8.6±0.6 | 62.8 | 2294.5±15.2 |
| | Back | 3009.1 ± 48.5 | | | |
| 20 mM | Front | 10.0 | 76.6±8.89 | 90.0 | 1999.8±9.8 |
| | Back | 2531.2 ± 31.2 | | | |
| 40 mM | Front | 8.6 | 80.4±10.1 | 91.4 | 2494.7±13.2 |
| | Back | 2844.9 ± 26 | | | |
| 60 mM | Front | 13.8 | 370.0±27.1 | 86.2 | 1483.8±12 |
| | Back | 61.8 | 11.2±0.2 | 38.2 | 1701.1±6.5 |
| 80 mM | Front | 2318 +11 | | | |
| | Back | 52.8 | 7.2 ± 0.2 | 47.2 | 2169 ± 12 |
| 100 mM | Front | 1115.5±2.8 | | | |
| | Back | 67.4 | 9.1±0.1 | 32.5 | 1344.7±4.7 |

Table S3 Biexponential fitting of the TRPL Traces (with annealing) in Fig. S4b and Fig. 3b.

| Concentration | | $A_1$ | $\tau_1$ | $A_2$ | $\tau_2$ |
|---|---|---|---|---|---|
| *Ctl* | Front | 37.2 | 40.9±2.8 | 62.8 | 787±8.7 |
| | Back | 37.6 | 118.9±6 | 62.4 | 118.9±6±62.4 |
| *10 mM* | Front | 38.3 | 12.9±1.2 | 61.7 | 572.6±6.1 |
| | Back | 37.8 | 14.3±0.9 | 62.2 | 14.3±0.9±62.2 |
| *20 mM* | Front | 44.4 | 21.9±1.3 | 55.6 | 911.6±8.3 |
| | Back | | 1136.8±14.8 | | |
| *40 mM* | Front | 69.3 | 6.1±0.2 | 30.7 | 1839.1±20.2 |
| | Back | | 1685.2±15.3 | | |
| *60 mM* | Front | 84.4 | 14.4±0.2 | 15.6 | 2397.5±37.9 |
| | Back | | 2906.9±38.9 | | |
| *80 mM* | Front | 88.2 | 14.6 ±0.2 | 11.8 | 1428 ± 19 |
| | Back | | 2786 ± 37 | | |
| *100 mM* | Front | 85.9 | 11.3±0.1 | 14.1 | 562.7±5.8 |
| | Back | | 2596.6±28.6 | | |

Table S4 Tabulated device parameters with variation of PTEAI concentration (10mM – 100mM), and with no annealing after the PTEAI treatment

| Concentration | PCE (%) | $V_{OC}$ (mV) | Jsc (mA cm$^{-2}$) | FF (%) |
| --- | --- | --- | --- | --- |
| *Ctl* | 15.97 | 1076 | 22.27 | 66.6 |
| *10mM* | 14.54 | 1011 | 23.53 | 61.12 |
| *20mM* | 18.6 | 1113 | 23.51 | 71.11 |
| *40mM* | 19.54 | 1128 | 23.31 | 74.31 |
| *60mM* | 16.15 | 1082 | 23.01 | 64.89 |
| *80mM* | 5.123 | 941 | 13.51 | 40.3 |
| *100mM* | 1.196 | 1045 | 2.325 | 49.24 |

Table S5 Tabulated device parameters with variation of PTEAI concentration (10mM – 100mM), and with annealing after the PTEAI treatment

|  | PCE (%) | $V_{OC}$ (mV) | $J_{SC}$ (mA cm$^{-2}$) | FF (%) |
| --- | --- | --- | --- | --- |
| *Ctl* | 15.28 | 1066 | 21.94 | 65.3 |
| *10mM* | 16.89 | 1083 | 21.91 | 71.14 |
| *20mM* | 5.271 | 907.6 | 20.13 | 28.86 |
| *40mM* | 0.4648 | 899.6 | 1.52 | 33.98 |
| *60mM* | 0.3663 | 914.2 | 1.364 | 29.37 |
| *80mM* | 0.6744 | 928.3 | 2.598 | 27.96 |
| *100mM* | 0.1055 | 906.4 | 0.3443 | 33.8 |